\documentclass{aa}
\usepackage{graphics}
\usepackage{epsfig}

\begin{document}

   \thesaurus{06     
              (02.13.3;  
               02.16.2;  
               03.20.2;  
               08.03.4;  
               08.09.2 S~Per;  
               08.13.1;  
               08.19.3)}  
   \title{Circular Polarization of Circumstellar Water Masers around S~Per}
   \titlerunning{Circular Polarization of Circumstellar Water Masers}


   \author{W. Vlemmings\inst{1}\and
        P.J. Diamond\inst{2}\and
        H.J. van Langevelde\inst{3}\
          }

   \offprints{WV (vlemming@strw.leidenuniv.nl)}

   \institute{Sterrewacht Leiden, Postbus 9513, 2300 RA Leiden, 
              the Netherlands
        \and
         Jodrell Bank Observatory, University of Manchester, Macclesfield,
                    Cheshire, SK11 9DL, England     
        \and
        Joint Institute for VLBI in Europe, Postbus 2, 
                7990~AA Dwingeloo, The Netherlands
        }

   \date{Received ; accepted }

\maketitle

\begin{abstract} 
We present the first circular polarization measurements of
circumstellar H$_{2}$O masers. Previously the magnetic field in
circumstellar envelopes has been estimated using polarization
observations of SiO and OH masers. SiO masers are probes of the high
temperature and density regime close to the central star. OH masers
are found at much lower densities and temperatures, generally much
further out in the circumstellar envelope. The detection of the
circular polarization of the (6$_{16}$--5$_{23}$) rotational
transition of the H$_{2}$O maser could be attributed to Zeeman
splitting due to the magnetic field in the intermediate temperature
and density regime. The fields inferred here agree well with predicted
values for a combination of the $r^{-2}$ dependence of a solar-type
magnetic field, and the coupling of the field to the high density
masing regions. We also discuss the unexpected narrowing of the
circular polarization spectrum.
\end{abstract} 

\maketitle

\section{Introduction}
\label{intro} 
 High mass loss in late type stars produces a circumstellar envelope
 (CSE) in which several different maser species can be found. These
 masers, especially SiO, H$_{2}$O and OH, are excellent tracers of the
 dynamics and kinematics of the CSEs. Polarization observations of
 these masers have revealed the strength and structure of the magnetic
 field throughout the CSE. Observations of SiO maser polarization have
 shown highly ordered magnetic fields 
 close to the central star, at radii of 5-10 AU where the SiO maser
 emission occurs (Kemball \& Diamond, 1997). The standard Zeeman
 interpretation gives magnetic field strengths of $\approx$ 5-10
 G. However, a non-Zeeman interpretation has been proposed by Wiebe \&
 Watson (1998), which only requires fields of $\approx$~30 mG. At much
 lower densities and temperatures and generally much further from the
 star, OH maser observations measure fields of $\approx 1$ mG
 (Szymczak \& Cohen, 1997; Masheder et al., 1999). But for the
 intermediate region, at distances of a few hundred AU, no information
 is available. In this region the H$_{2}$O maser emission
 occurs. Since water is a non-paramagnetic molecule, determination of
 the magnetic field is significantly more difficult. The Zeeman
 splitting of H$_{2}$O is extremely small for the field strengths
 expected (few hundred mG), only $\approx$ $10^{-3}$ times the typical
 Gaussian line width of the H$_{2}$O maser line ($\Delta\nu_{\rm L}
 \approx 20$ kHz). However, Fiebig \& G\"usten (1989, hereafter FG)
 showed that in the presence of such magnetic fields the Zeeman
 splitting can be detected with high spectral resolution polarization
 observations. Their observations targeted strong interstellar
 H$_{2}$O maser features. The observations presented here give the
 first results of circular polarization measurements of the H$_{2}$O
 masers found in CSEs. We have used a method similar to that used in
 FG to determine the magnetic field strength parallel to the line of
 sight. Like FG we are limiting ourselves to a Zeeman interpretation
 of the observed splitting, although a non-Zeeman interpretation has
 been presented in Neduloha \& Watson (1990).

\section{Observations}
\label{obs}

The observations were performed at the NRAO\footnote{The National
  Radio Astronomy Observatory is a facility of the National Science
  Foundation operated under cooperative agreement by Associated
  Universities, Inc.}  Very Long baseline Array (VLBA) in December
  1998. We observed 4 late type stars (S Per, U Her, VY CMa and NML
  Cyg), the results presented here are the first results for the
  supergiant S Per, the source with strong water maser features and
  relatively simple structure. The stellar velocity V$_{\rm LSR}$ of S
  Per is $-38.5$ km~s$^{-1}$.The beam width at 22.235 GHz, the
  frequency of the $6_{16}-5_{23}$ rotational transition of H$_{2}$O,
  is $\approx 0.7 \times 0.3$ mas. This allows us to resolve the
  different H$_{2}$O maser features in the CSE.  The data were
  correlated twice, once with modest ($7.8$ kHz $= 0.1$ km~s$^{-1}$)
  spectral resolution, which enabled us to generate all 4 polarization
  combinations (RR, LL, RL and LR). The second correlator run was
  performed with high spectral resolution ($1.95$ kHz $= 0.027$
  km~s$^{-1}$), necessary to detect the circular polarization
  signature of the H$_{2}$O Zeeman splitting, and therefore only
  contained the two polarization combinations RR and LL. The data
  produced by the first correlator run were used to accurately
  calibrate the R- and L-polarization. The calibration solutions were
  obtained on R and applied to L after we determined the relative R-L
  corrections, assuming the R- and L-polarization line profiles to be
  similar. The solutions were then applied to the data-set produced by
  the second correlator run, which we used to determine the circular
  polarization V for the separate maser features. The data analysis
  followed the method of Kemball, Diamond \& Cotton (1995).

\section{Background}
\label{background}

The H$_{2}$O ($6_{16}-5_{23}$) rotational transition consists of 6
hyperfine components. Analyses of interstellar water masers indicate
that all 6 hyperfine components contribute to the maser (Walker,
1984). The observed intensity ratios however, frequently deviate from
those obtained from molecular transition probabilities (Moran et al.,
1973; Genzel et al., 1979). Our analysis is performed using fitted
line ratios for the 3 strongest hyperfine components ($F=7-6, 6-5$ and
$5-4$). The separation between these components is $0.45$ and $0.58$
km~s$^{-1}$ respectively. The weakest components ($F=5-6, 5-5$ and
$6-6$) are separated by more than $2.5$ km~s$^{-1}$. They are not
observed in the total power spectrum so we have not included them in
our analysis. Our treatment follows closely the analysis performed in FG.
Here we have added the possibility of multiple masering
hyperfine components. We assume the strongest hyperfine component
($F=7-6$) to be the dominant transition. Because of the
non-paramagnetic nature of the H$_{2}$O molecule, the Zeeman splitting
is extremely small. It is due to the interaction of the nuclear
magnetic moment with the external B field. Thus the splitting is a
factor of $10^3$ weaker than that for radicals like OH.  In the weak
field limit, the split energy $\Delta E_{\rm Z}$ of a given energy level
(J,F,I) is determined by:
\begin{eqnarray}
\Delta E_{\rm Z} &=& - \{\alpha_{J}g_{J} +
\alpha_{I}g_{I}\}\cdot\mu_{N}M_{F}\cdot B_{\rm Gauss} \\
\label{eq1}
\rm{with:}\nonumber\\
\alpha_{J} &=& \{J(J+1)+F(F+1)-I(I+1)\}/2F(F+1),\nonumber\\
\alpha_{I} &=& \{F(F+1)+I(I+1)-J(J+1)\}/2F(F+1).\nonumber
\end{eqnarray}
This corresponds to a characteristic frequency shift of the order of
$\Delta \nu_{\rm Z} \approx 10^3$ Hz$\cdot [B_{\rm Gauss}]$. Where $I(=1)$
is the nuclear spin, $\mu_{N}$ the nuclear magneton and $M_F$ the
magnetic quantum number; $g_I = 5.585$, and the $g_J$-factors $g_6 =
0.6565$ and $g_5 = 0.6959$ are from Kukolich (1969).

\begin{figure} 
 \hspace{0.3cm}
\psfig{figure=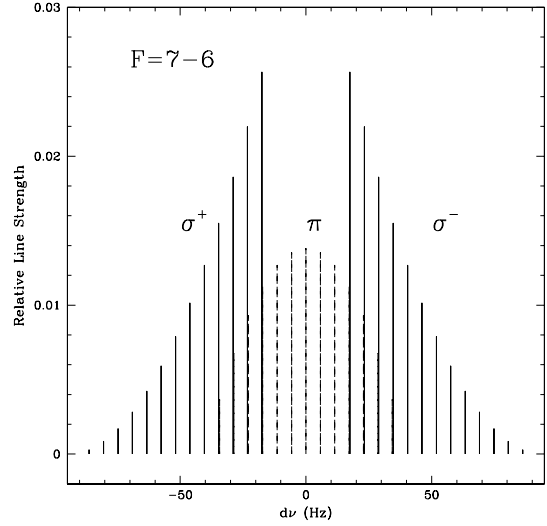,width=0.42\textwidth}
   \hfill
   \caption{The Zeeman pattern of the $F=7-6$ hyperfine component for
   an external field of 50 mG.}  \label{zm}
\end{figure}

\begin{figure}
 \hspace{0.3cm}
\psfig{figure=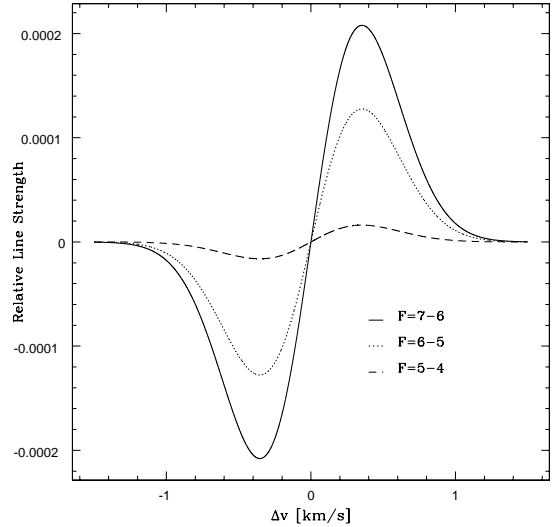,width=0.42\textwidth}
   \hfill \caption{Synthetic V-spectra for $F=7-6, 6-5$ and $5-4$
   calculated for an external field of 50 mG and a line width of
   $\Delta v_{\rm L}=0.5$ km~s$^{-1}$.} \label{synth}
\end{figure}

Each hyperfine component will split into 3 groups of lines
($\sigma^{+}, \sigma^{-}$ and $\pi$), as seen in Fig.~\ref{zm} for one
of the hyperfine components.  The relative strengths of the transition
probabilities have been determined by Deguchi \& Watson (1986). For a
magnetic field B parallel to the line of sight the Zeeman pattern
consists of the two circular polarized $\sigma$ components only. The
right- and left-handed (RR and LL) spectra, corresponding to the
$\sigma^{\pm}$ components will only be slightly shifted against each
other $(\Delta v_{\rm Z} \approx 10^{-3~ \rm{to}~-4}$ times $\Delta
v_{\rm L}$).  As a result, the observed V-spectrum (RR-LL) will be a
sine-shaped function, corresponding to the derivative $I'$ of the
total power spectrum. The amplitude of this function depends on the
maser line width, the magnetic field strength, and on which hyperfine
components actually contribute to the maser. By calculating synthetic
V-spectra from the Zeeman pattern for different line widths, magnetic
field strength and hyperfine combinations we find the following
relation for the percentage of circular polarization:
\begin{equation}
P_{\rm V} \equiv (V_{\rm max} - V_{\rm min})/I_{\rm max}
 = A_{\rm F-F'}\cdot B_{\rm Gauss}/\Delta v_{\rm L}. 
\label{eq2}
\end{equation}
Here $V_{\rm max}$ and $V_{\rm min}$ are the maximum and minimum of
the synthetic V-spectrum fitted to the observations. $\Delta v_{\rm L}
[\rm{km~s}^{-1}]$ is
the Gaussian line width of the total power spectrum, and $I_{\rm max}$
is the peak flux of the maser feature. $B$ is the magnetic field
component along the line of sight.  The $A_{F-F'} [\cdot 10^{-3}]$
coefficient depends on the masering hyperfine components. The
$A_{F-F'}$ coefficients have been determined by calculating 
$P_{\rm V}$ from synthetic V-spectra, determined for a
series of magnetic field strengths $B$, and for the different hyperfine
components. For the $F=7-6, 6-5$ and $5-4$ components individually we
find $A_{F-F'} = 16.22, 10.00$ and $1.23$ respectively. For a fitted
combination of hyperfine components we find slightly different values.
An example of synthetic V-spectra for the three hyperfine lines is
shown in Fig.~\ref{synth}.

However, due to the complex interactions between the various hyperfine
components in the maser regime, deviations from the V-spectrum
proportionality are possible. Detailed radiative transfer treatment,
as performed by Nedoluha \& Watson (1992; hereafter NW), for instance,
resulted in $A_{F-F'} = 23.9$.

\section{Results}
\label{res}

\begin{figure} 
 \hspace{0.3cm}
\psfig{figure=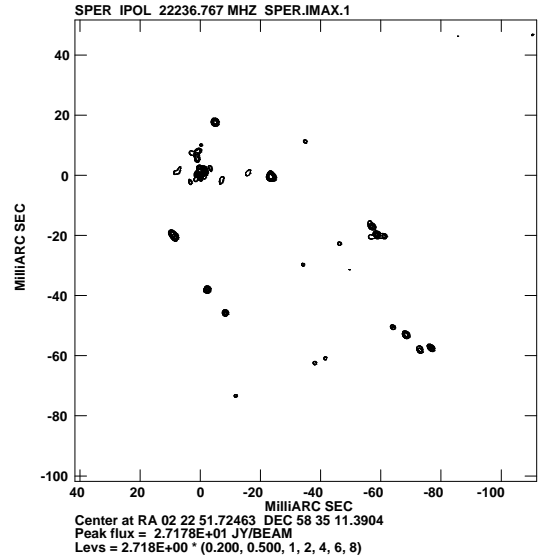,width=0.42\textwidth,angle=-90}
   \hfill \caption{Total intensity image of the H$_2$O maser features
   around S~Per.}  \label{sper}
\end{figure}

\begin{figure} 
 \hspace{0.3cm}
\psfig{figure=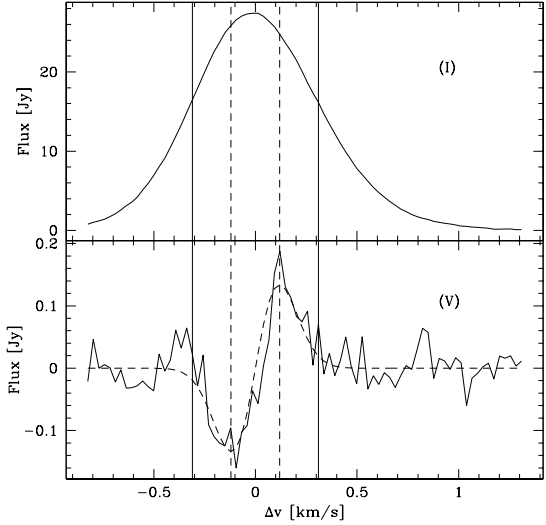,width=0.42\textwidth}
   \hfill \caption{Total power ($I$) and circular polarization ($V$)
   spectrum of the brightest H$_{2}$O maser feature around S~Per. The
   dashed line is the fit of the synthetic $V$-spectrum to the
   observed spectrum. Also shown are the observed (dashed) and
   expected (solid) positions of the minimum and maximum of the
   V-spectrum.}  \label{V}
\end{figure}

Fig.~\ref{sper} shows the total intensity map of the water maser
features surrounding S~Per. We are able to identify most of the maser
features detected in earlier observations (Diamond et al., 1987;
Marvel, 1997). The positions are relative to the brightest maser
feature, for which we have managed to determine the circular
polarization spectrum shown in Fig~\ref{V}. This figure also shows a
$\chi^2$-fit to the sine-shape spectrum. The amplitude of the
V-spectrum is only a small fraction ($\approx 1\%$) of the total
power, so we have only been able to determine the Zeeman splitting,
and thus magnetic field strength, for the brightest maser feature.  We
do not detect circular polarization in any of the other bright maser
features although, if it was present at the same absolute level as in the
brightest feature, we would have detected it. This further confirms
our detection, because a scaled down version of the total power $I$
would also have been detectable on the other strong features if
calibration errors were significant.

The total power spectrum indicates that one of the hyperfine
components clearly dominates, since the splitting of the hyperfine lines
should otherwise have been observable.  We have performed a fit to the
total power spectrum to determine the maser line width and the best
fitted ratio for the three strongest hyperfine transitions. Using this
line ratio we have calculated the $A_{F-F'}$ coefficient as described
above. In this case we find $A_{F-F'} = 15.54$. Using the fitted
Gaussian line width ($\Delta v_{\rm L}=0.44~\pm~0.01$ km~s$^{-1}$) and
$P_{\rm V} = (9.9 \pm 0.5) \cdot 10^{-3}$ in Eq.~\ref{eq2}, we find
for the magnetic field strength along the line of sight $B_{||} = 279
\pm 30$~mG. 
If only the $F=7-6, 6-5$ or $5-4$ hyperfine transition contributes the
magnetic field should be scaled by $0.96, 1.55$ or $12.63$
respectively. As seen in Fig.~\ref{V} the V-spectrum is negative on
the blue shifted side, therefore the observed component $B_{||}$ is
pointing away from us. This result is the first measured circular
polarization in the circumstellar water maser region.

\section{Discussion}
\label{disc}

 As discussed above, the magnetic field strength was obtained by using
a best fitted line ratio for the three main hyperfine components. A
radiative transfer treatment for the polarized maser radiation
of the $6_{16} - 5_{23}$ H$_{2}$O rotational transition was performed
by NW. Some difficulties exist matching their results to our
observations of the total intensity spectrum, which shows an almost
Gaussian shape. Their treatment also did not predict the
anti-symmetric shape of the V-spectrum as shown in our observations
and those by FG. However, since the calibration performed here and by
FG assumes similar R- and L-polarization line profiles, the
anti-symmetric shape is a necessary result of the treatment of the
data.

 The observed V-spectrum however, is not directly proportional to $I'$
either, as would have been expected in the simple model. The minimum
and maximum of the sine-shaped function are not located at $\pm
\sigma/\sqrt{2}$, with $\sigma$ being the observed Gaussian line
width. Instead they are found at $\Delta\nu$ smaller by a factor of
$\approx 2.5$ with respect to the line width fitted to the total power
spectrum. Possibly, this narrowing of the V-spectrum can be attributed
to the overlap of the multiple hyperfine components, as predicted by
the treatment and analysis of NW. The observed effect, however, still
seems too large. Due to this our magnetic field strength could be
overestimated by at most a factor of 2. The interstellar water masers
observed by FG did not show this narrowing.

Because of the narrowing of the V-spectrum it is difficult to address
the saturation state of the maser. Elitzur (1998) showed that the
observed circular polarization spectrum would be a good indication of
the maser saturation state. If the ratio of $V(\nu)/I'(\nu)$ increased
towards the wings of the line instead of showing a constant ratio the
maser is thought to be unsaturated. This is however, almost the
opposite effect of what we observe. Until the narrowing of the
circular polarization is fully explained our observations are difficult to
reconcile with a specific saturation state.

We have also tried to determine the linear polarization of the maser
features around S~Per, but no indication of linear polarization has
been found. This is consistent with the observations of Barvainis
\& Deguchi (1989). They explain that the absence of linear
polarization is probably due to the fact that masers are not very
strongly saturated, and that infrared radiation does not contribute
significantly to the pumping process.

The magnetic field strength derived this way is within the range
estimated by previous observations of SiO and OH masers. As noted
before, polarization observations of SiO masers close to the central
star reveal fields of 5-10 G, assuming a standard Zeeman
interpretation. OH maser observations of features around S~Per
indicate a field of slightly less than 1 mG (Masheder et al.,
1999). Based on these values, the dependence of $B \propto r^{-2}$ for
a solar-type magnetic field is the most likely. For a dipole medium,
the magnetic field is expected to vary with $r^{-3}$, which appears to
be too steep to accurately describe the observations.

 In S~Per, the H$_{2}$O and OH masers are observed to exist at similar
{\it projected} distances (Masheder et al., 1999). This would disagree with
the observed differences in the magnetic field strengths, except if
the magnetic field could remain frozen in high density clumps.
The magnetic field strength is then expected to vary with number
density as $B \propto n^k$, with $1/3 \le k \le 1/2$ from theoretical
predictions (e.g Mouschovias 1987), where $n$ is the number
density. SiO masers are observed in high density clumps at $5-10$ AU
from the central star. H$_{2}$O masers exist in similar clumps at
distances of a few hundred AU, with the magnetic field lines frozen in
the dense medium. Richards et al. (1999) show that the OH and H$_2$O
maser clumps avoid each other, although located at similar projected
distances. They conclude that the density ratio between the H$_2$O
maser clumps and the OH in the surrounding medium only needs to be a
factor of 50. However, magnetic fields frozen into the maser clumps
would require a density ratio of $\approx 10^4$ to explain the
difference in field . This seems to indicate that actual coexistence
between the OH and H$_2$O masers is unlikely.

In conclusion, although the exact influence of the hyperfine
interaction is not yet clear, we derive a magnetic field
strength of $B_{||} = 279 \pm 30$ mG.

{\it Acknowledgments:} 
This project is supported by NWO grant 614-21-007.

\end{document}